# An Investigation of Intra-Urban Mobility Pattern of Taxi Passengers: Multi-Case of Three Cities

Ling Zhang[1], Shuangling Luo[2], Haoxiang Xia[1]

1. Institute of Systems Engineering, Dalian University of Technology, Dalian 116024 China

2. College of Transportation Management and Centre for Collaborative Innovations on Comprehensive Transportation, Dalian Maritime University, Dalian 116026, China

## Abstract

The study of human mobility patterns is of both theoretical and practical values in many aspects. For long-distance travels, a few research endeavors have shown that the displacements of human travels follow the power-law distribution. However, controversies remain in the issue of the scaling law of human mobility in intra-urban areas. In this work we focus on the mobility pattern of taxi passengers by examining five datasets of the three metropolitans of New York, Dalian and Nanjing. Through statistical analysis, we find that the lognormal distribution with a power-law tail can best approximate both the displacement and the duration time of taxi trips, as well as the vacant time of taxicabs, in all the examined cities. The universality of scaling law of human mobility is subsequently discussed, in accordance with the data analytics.

## 1. Introduction

The study of human mobility patterns is of great importance in many aspects, such as urban planning [1], control of epidemic spreading [2], tourism management [3], emergency management [4] and transport prediction and planning [5]. However, mainly due to the limitations in collecting and analyzing the large-scale mobility data, the mobility or travel patterns of massive populations in different geographical scales had not been well-studied until recently.

In the last decade, with the increasing availability of different types of large-scale location-based data of massive populations, as well as the rapid prominence of data-driven "computational social science" [6], our understandings of human mobility patterns have been greatly deepened. This subject research has also attracted great attention of scientists in the areas of non-linear physics and complex systems science. One well-noted work is given by Brockmann et al. [7]. By tracing the circulation of bank notes in the United States of America, they identified that the travelling distances of banknote carriers followed a power-law distribution, indicating that the carriers' travels are alike to Lévy flights with attenuation of dispersal. A two-parameter continuous-time random walk model was then developed to reproduce the observed travelling pattern in their work. González et al. [8] studied the trajectories of 100,000 mobile-phone users, finding that the distribution of displacements over all users was well approximated by a truncated power-law. González et al.'s work was subsequently extended by Song et al. [9], who developed an individual-mobility model based on the idea of preferential return to account for the scaling law observed in mobile-phone trajectories. Another interesting model was proposed by Han et al. [10], in which the power-law of travel distances is explained by the hierarchical structure of transport systems.

The aforementioned series of studies are fascinating. However, datasets used in these studies are with limited accuracy to depict human mobility in smaller geographical scales. For the dataset of banknote circulation, one recorded travel of a banknote does not precisely reflect the movement of any single person, since the banknote may change hands several times during the period of this recorded banknote travel. Especially, it would be difficult to trace the short-distance travels of human beings through the banknote records, e.g. within the travel range of less than 10km as shown in Brockmann et al.'s own paper. The inaccuracy of the mobile-phone dataset lies in its shortage of tracing short-distance travels, since the mobile-phone positioning records depend on the locations of communication company's base stations and the travels within the range of one single base-station cannot be properly traced. Due to such limitations, a question can naturally be raised whether the human travels in shorter-distances, especially within the range of intra-urban areas, statistically follow similar mobility patterns as observed in the longer-distance (i.e. from around ten kilometers to thousands of kilometers) datasets. Comparing with the datasets of banknotes and mobile phones, the data of Global-Positioning-System (GPS) trajectories of vehicles reflect human mobility more directly and with finer granularity. The availability of vehicle GPS trajectory data stimulates various studies on the mobility patterns in intra-urban areas.

Based on the GPS data generated by fifty taxicabs during a six-month period in Sweden, Jiang et al. [11] found that the travel distances by taxi passengers follow a two-phase power-law distribution. The power-law is also supported by Yao and Lin [12], according to their analysis of taxi trajectories in a South China city. In comparison, various other studies on the datasets of private cars [13] and taxis [14, 15] showed the exponential distributions of travel distances. A gamma distribution of travel distances was suggested by Veloso and Phithakkitnukoon [16], who investigated the taxi data during the course of five months in Lisbon, Portugal. Roth et al. [17] analyzed the dataset of individuals' subway travels in London during one week, finding that the travel distances approximate a negatively binomial distribution. The previous researches indicate that the intra-urban travels do not simply follow the same power-law of longer-distance travels. What's more, the distributions of intra-urban travels fitted in different researches contradict with one another. Such situation demands more thorough investigations based on intra-urban mobility data that are with better temporal and geographical coverage, in order to examine whether there is a universal scaling law for human travels in intra-urban areas.

As an attempt that continues the preceding endeavors on intra-urban mobility, in this paper we examine the travel distributions of taxi passengers by analyzing five datasets of taxi trajectories in the cities of Dalian, Nanjing and New York. The inconsistent results obtained in previous researches imply that the human mobility patterns may be inherently intricate. Therefore, in this work we limit our research scope in order to improve the accuracy of results. On one hand, we restrict the travel mode to the single means of taxi-taking. On the other, the travel distances are limited to the range of a single city, more specifically within the metropolitan areas as the examined cities are all with more than one million in population although these cities are significantly different in their geographical characteristics and traffic situations.

The remainder of this paper is organized as follows. In Section 2, the datasets to be used are described and the method of data cleaning is introduced. We describe the analysis method in Section 3. In Section 4, three statistical metrics are respectively analyzed, namely the distributions of travel displacements, the distributions of duration time of travels, and the distributions of vacant time of taxicabs. Finally, the results are discussed and the whole paper is concluded in Section 5.

## 2. Data Description and Cleaning

In order to more precisely explore human mobility patterns in the intra-urban areas, we use five datasets of GPS trajectories of taxis. The dataset $D_1$ was collected from 7804 taxis within the urban areas in Dalian, China, from Jan. 1st to May. 28th, 2014. This dataset is from the Department of Road Transport of Dalian City, covering the trajectory data of most taxicabs of that city, i.e. 7804 out of about 8900 taxicabs. The datasets $D_2$, $D_3$ and $D_4$ are from the open datasets by New York City (NYC) Taxi and Limousine Commission, which can be publicly accessed at the Website[1]. The dataset $D_2$ covers the trajectories of more than 13,000 yellow taxicabs in November and December of 2015. Both $D_3$ and $D_4$ are from NYC green taxicabs in whole year of 2014 ($D_3$) and 2015 ($D_4$). The dataset ($D_5$) records the GPS trajectories of 7700 taxis in the city of Nanjing, China. In all the five datasets, the following information is contained, i.e., vehicle ID (anonymized in $D_2$, $D_3$ & $D_4$), pick-up time, drop-off time, travel distance, pick-up longitude, pick-up latitude, drop-off longitude and drop-off latitude. Three main statistical metrics are extracted from the records, i.e., the passenger's trip displacement, the passenger's trip duration time, and the vacant time of taxicab. The basic information of the five datasets is summarized in Table 1.

Table 1 Basic information of the five datasets

| Dataset | $D_1$ | $D_2$ | $D_3$ | $D_4$ | $D_5$ |
| --- | --- | --- | --- | --- | --- |
| Dataset Name | Dalian | NYC yellow | NYC green | NYC green | Nanjing |
| Effective Days | 148 | 61 | 365 | 365 | 15 |
| # of Taxicabs | 7804 | >13000 | >6000 | >12000 | 7700 |
| # of Trips per day | 193k | 485k | 43k | 60k | 140k |
| Avg. Displacement(km) | 6.335 | 3.092 | 3.055 | 2.951 | 4.243 |
| Avg. Duration(min) | 15 | 15 | 13 | 13 | 17 |

To note: as the vehicleIDs are protected in NYC datasets, we don't have the exact number of taxicabs.

Before analyzing these datasets, data cleaning is processed to eliminate the noise data. First, records with unusually-short displacements (i.e. less than 0.5km) are removed, as walk is the more common choice for a less-than-0.5km trip. Hence, we consider those recorded less-than-0.5km trips as error data generated by incorrect GPS traces. Second, we discard the trips that are farther than 100 km (50 km for NYC datasets), as the accounts of such long-distance travels are very few in the datasets. Finally, the trips with less-than-one-minute and more-than-3-hours in duration are also ignored, since such trips are unusual for an average taxi passenger.

## 3. Method

### 3.1 Cullen and Frey Graph

In this work we use descriptive statistics to select candidate theoretic distributions or models to fit the aforementioned trajectory data. In descriptive statistical analysis, kurtosis and skewness are two widely-used coefficients. Kurtosis is a measure to depict the tailedness of a distribution in comparison with the normal distribution for which the kurtosis is equal to 3. Distributions with kurtosis greater than 3 are called "leptokurtic", while distributions with kurtosis less than 3 are called "platykurtic". Skewness measures the level of asymmetry of a distribution on its mean. If the skewness of a distribution is positive, the part of its probability density function that is on the right of the mean is "fatter" than the left part. On the contrary, the negative skewness of a distribution reveals that the left part of the density function is "fatter" than the right part.

---

[1] http://www.nyc.gov/html/tlc/html/about/trip_record_data.shtml.

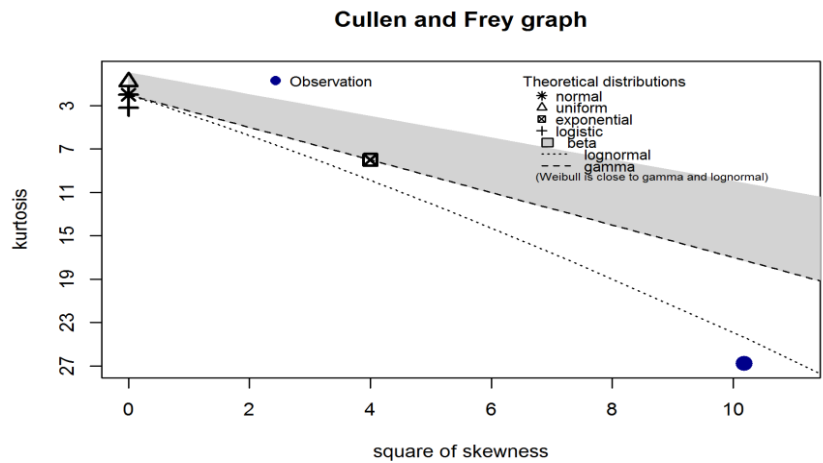

Figure 1 Cullen and Frey graph [18]

Combining the two coefficients, Cullen and Frey [18] used the skewness-kurtosis plot (i.e. Cullen and Frey Graph) to depict the skewness and kurtosis of the commonly-used distributions, as shown in Fig.1. For normal, uniform, exponential and logistic distributions, the values of skewness and kurtosis are unique. Thus each of those distributions is represented as a single point in the plot. Gamma and lognormal distributions are represented as lines, while the Beta distribution is represented by a shaded area in the skewness-kurtosis plot.

Table 2 The skewness and kurtosis of the three statistics

| Statistic | Datasets | Skewness | Kurtosis |
|---|---|---|---|
| Displacement | $D_1$ | 3.19 | 27.78 |
|  | $D_2$ | 2.79 | 11.74 |
|  | $D_3$ | 2.43 | 11.88 |
|  | $D_4$ | 2.58 | 13.11 |
|  | $D_5$ | 4.74 | 41.18 |
| Duration | $D_1$ | 3.56 | 28.94 |
|  | $D_2$ | 2.30 | 11.833 |
|  | $D_3$ | 2.17 | 12.18 |
|  | $D_4$ | 2.49 | 16.11 |
|  | $D_5$ | 2.95 | 19.61 |
| Vacant time | $D_1$ | 3.49 | 19.06 |
|  | $D_5$ | 3.43 | 17.37 |

We calculate the values of kurtosis and skewness of the records of passenger displacement, passenger duration time and vacant time of taxicab in the five datasets, as shown in Table 2. For all the data series, the kurtosis and skewness values are greater than 2, indicating that these data series would probably be well-fit by some right-skewed and leptokurtic distribution, whilst normal, uniform and exponential distributions may not fit the data well. Therefore, subsequently we select three major right-skewed and leptokurtic distributions as candidates to fit the data, namely lognormal, Weibull, and gamma distributions. However, as some previous researches have illustrated the good-fit of the exponential distribution and the power-law distribution, these two type of distributions are also taken into account in the following analysis.

### 3.2 Selection of candidate distributions

As mentioned above, lognormal, Weibull, Gamma, exponential, and power-law distributions are selected as candidates to fit our datasets. Their probability density functions are shown below.

The probability density function of lognormal distribution is defined by Eq. (1)

$$P(x) = \frac{1}{x\sigma\sqrt{2\pi}} exp\left(-\frac{(lnx-\mu)^2}{2\sigma^2}\right) \quad (1)$$

where μ is the mean and σ is the standard deviation.

The Weibull distribution is defined by the probability density function of Eq. (2).

$$P(x) = \frac{\alpha}{\beta}\left(\frac{x}{\beta}\right)^{\alpha-1} exp\left(-\left(\frac{x}{\beta}\right)^{\alpha}\right) \quad (2)$$

where α > 0 is the shape parameter, and β > 0 is the scale parameter.

The probability density function of Gamma distribution is defined by Eq. (3).

$$P(x) = \frac{\beta^{\alpha}}{\Gamma(\alpha)} x^{\alpha-1} exp(-\beta x) \quad (3)$$

where α and β, respectively, are shape and rate parameters.

The probability density function of exponential distribution is defined by Eq. (4).

$$P(x) = exp(-\lambda x) \quad (4)$$

where λ is the rate parameter.

Finally, the probability density function shown in Eq. (5) is used to define the power-law distribution.

$$P(x) = \frac{\alpha-1}{x_{min}}\left(\frac{x}{x_{min}}\right)^{-\alpha} \quad (5)$$

where $\frac{\alpha-1}{x_{min}}$ is the normalizing constant and α is the power parameter.

To find out the distribution that best fit the data from the candidate distributions, Akaike Information Criterion (AIC) [19] and Bayesian Information Criterion (BIC) [20] are two commonly-used fitting criteria. In this paper, Akaike information criterion or AIC is used as the criterion for measuring the goodness-of-fit between the theoretical distribution (i.e. the model) and empirical data. The overall procedure of model selection is comprised of three steps, as shown below.

a. Calculating the model parameters. In the paper, the parameters of candidate models are computed by Maximum Likelihood Estimation (MLE) [21].
b. Calculating the AIC score of each model. For the candidate model $i$ ($i\epsilon\{1,2,3,4,5\}$), the corresponding AIC score is computed by $AIC_i = -2logL_i + 2K_i$, where $K_i$ represents the number of parameters of model $i$, and $L_i$ is the likelihood function.
c. Selecting the best-fit model. AIC weights ($w_i$) [22] denotes the relative likelihoods of the model, and we use it as the criterion to select the best-fitting model. Let $AIC_{min} = min\{AIC_i\}$ and $\Delta_i = AIC_i - AIC_{min}$, and the AIC weights could be calculated by $w_i = \frac{exp(-\Delta_i/2)}{\sum_{j=1}^{m} exp(-\Delta_j/2)}$. The candidate model with largest AIC weight is the one fitting the actual data best.

## 4. Results and Analyses

### 4.1 Displacement distribution

By utilizing the previous datasets, we first examine the distributions of displacement, i.e. the distances of travel.

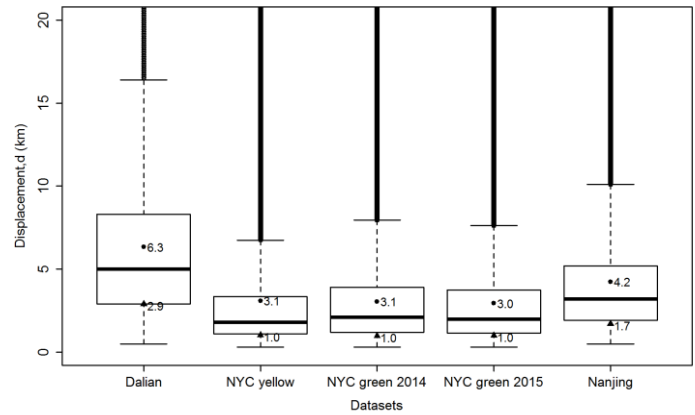

Figure 2 The box plots of displacement in the five datasets

Fig. 2 illustrates the box plots of the displacements extracted from the five datasets, where the black dots represent the mean values of displacements, and black triangle points denote the most frequent displacements. From Fig. 2, we can firstly find that in all plots the mean values are greater than the medians, indicating that all data series are comprised of abundant of short-distance travels and relatively-fewer long-distance travels. Among the five displacement data series, the average travel distance of taxi trips in NYC is approximately 3.0 km while that in Dalian is 6.3 km and that in Nanjing is 4.2 km. The median displacement value of trips in NYC is about 2.0 km, while that in Dalian is 5.0 km and that in Nanjing is 3.2 km. Moreover, the most frequent displacement in NYC is 1.0 km, which is also shorter than that in Dalian (2.9km) and in Nanjing (1.7km). The most frequent displacements in all three cities are shorter than the corresponding mean and median values of displacements. What's more, the shorter mean, median and most-frequent displacements in NYC than those in Nanjing and Dalian may possibly reflect the faster pace of life in NYC.

The previous plots reveal that the distributions of displacement are skewed in all three five datasets. Subsequently we examine the fitness of the five candidate theoretic distributions to the actual data, as shown in Figure 3.

Figure 3 illustrates the fittings of four datasets, namely, $D_1$ (Dalian), $D_2$ (New York Yellow), $D_4$ (New York Green 2015) and $D_5$ (Nanjing). As it is very similar with that of $D_4$, the fitting result of dataset $D_3$ is not illustrated in Figure 3 so as to ease the layout of the plots.

In Figure 3, the actual data points are shown as black empty circles, while the fitting data generated by the lognormal, Weibull, gamma and exponential distributions are respectively presented as red, blue, green and purple curves. By comparing the actual data points with the theoretical curves, we can intuitively estimate that the lognormal and exponential distribution fit the actual data better than the other two distributions. As for the power-law distribution, our observation is that this type of distribution does not generate good fit to the full-scoped data in general. However, the actual displacements that are greater than 18km can be well-fitted by the power-law, as illustrated by the orange lines in Fig.3. For such long-distance displacements, the power-law fit is apparently better than the other four types of distributions.

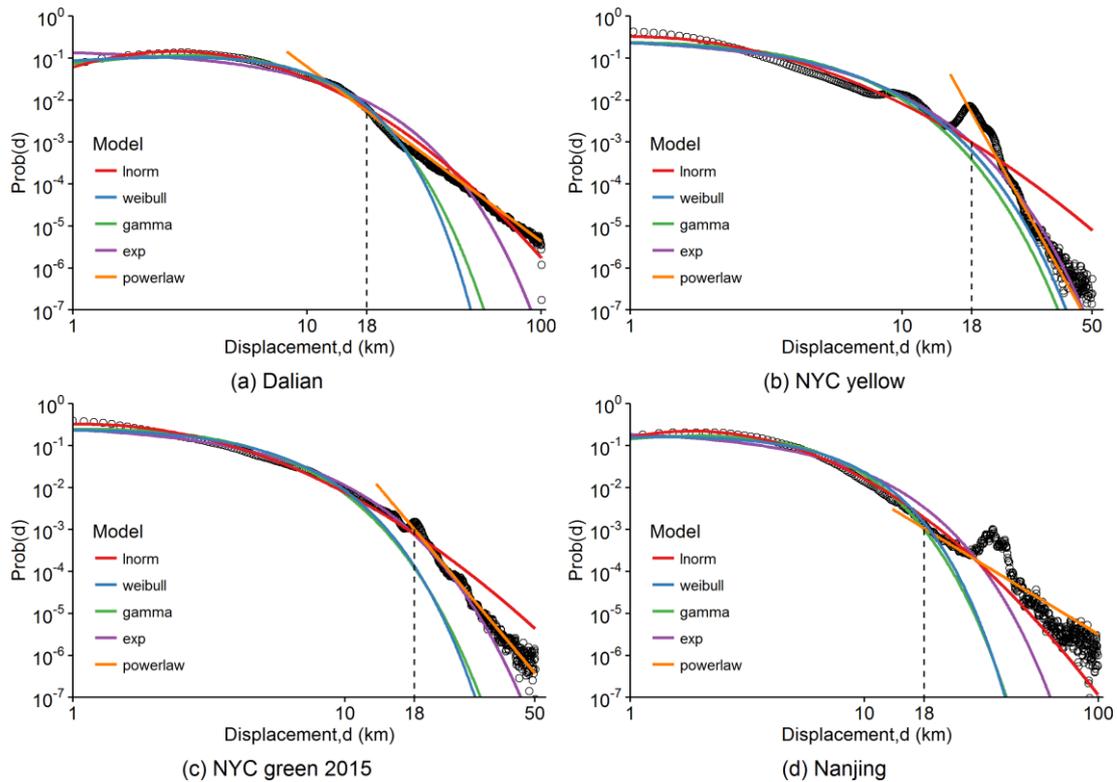

Figure 3 Fittings of the distributions of displacement at log-log scale

Furthermore, we calculate the AIC weights of the five candidate models, as shown in Table 3. It can be seen that the AIC weights of lognormal model are 1 in all datasets, while the AIC weights of the other four models are zero. This indicates that all the datasets are best fitted by the lognormal.

Table 3 The AIC weights of displacement

| Datasets | AIC weight | | | | |
| --- | --- | --- | --- | --- | --- |
| | Lognormal | Weibull | Gamma | Exponential | Power-law |
| $D_1$ | 1 | 0 | 0 | 0 | 0 |
| $D_2$ | 1 | 0 | 0 | 0 | 0 |
| $D_3$ | 1 | 0 | 0 | 0 | 0 |
| $D_4$ | 1 | 0 | 0 | 0 | 0 |
| $D_5$ | 1 | 0 | 0 | 0 | 0 |

The previous analysis indicates that the displacements in all five datasets can best be fitted by the lognormal distributions with a power-law tail for the long-distance travels. It is worth noting that the taxi trips that are within 18km respectively cover 97.4%, 98.5%, 99.95%, 99.98%, 98.5% of the recorded trips in the five datasets. Thus, the power-law tails are to some extent negligible in describing the overall statistics of the intra-urban displacements of the taxi passengers. By omitting the data records of long-distance travels that are best fitted by the power-law, Fig.4 more specifically shows the fitting results of lognormal, Weibull, gamma and exponential distributions to the displacements within the range of 18km, where the advantage of the lognormal distribution over the other three types of distributions is more apparent than in Fig. 3.

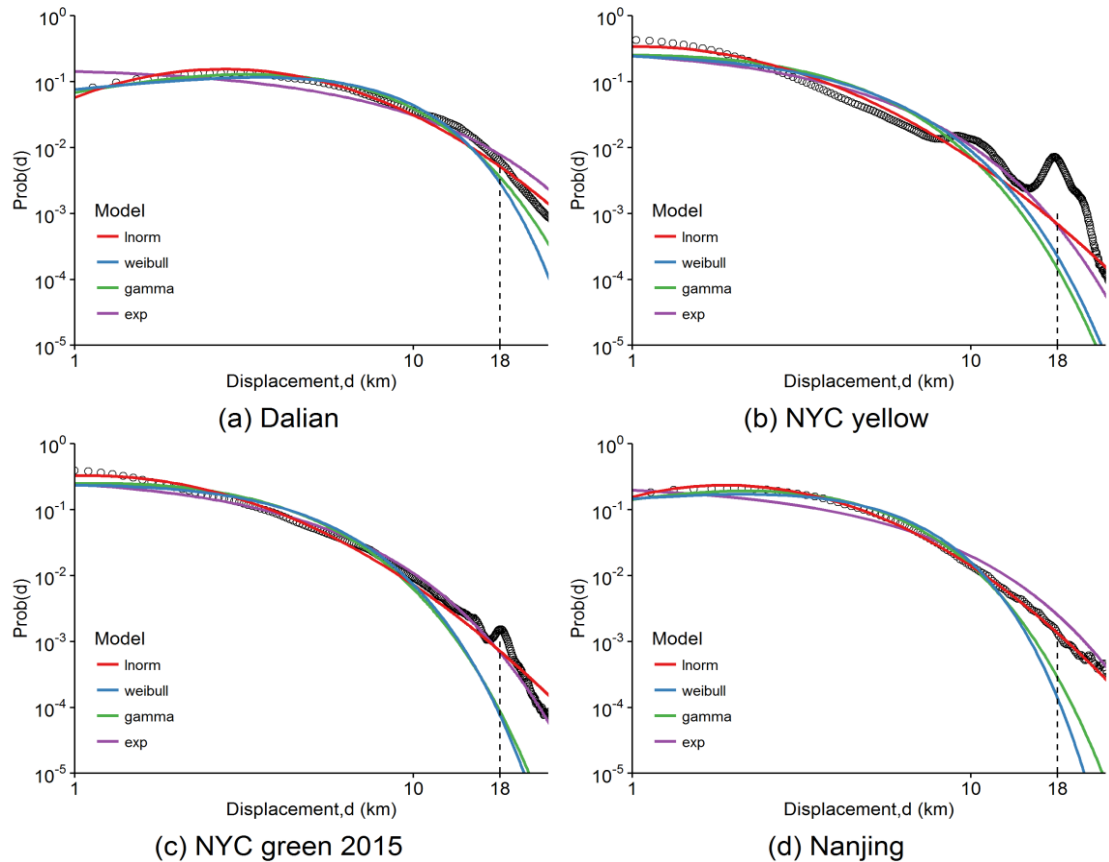

Figure 4 Fittings of the distributions of displacement within 18 km at log-log scale

In the previous fittings of the actual data records, the parameters of models are computed by the Maximum Likelihood Estimation (MLE) method. Table 4 lists the adopted parameters of the lognormal distributions to fit the displacements in the five datasets, and the 95% confidence bounds of lognormal parameters are listed in the brackets.

Table 4 Parameters for the lognormal distributions to fit the displacement data

| Datasets | MLE for lognormal parameters (with 95% CI bounds) | |
|---|---|---|
| | Mean ($\mu$ in Eq.(1)) | Standard Deviation ($\sigma$ in Eq. (1)) |
| $D_1$ | 1.573(1.5727,1.5733) | 0.758(0.7578,0.7582) |
| $D_2$ | 0.716(0.7154,0.7162) | 0.850(0.8498,0.8503) |
| $D_3$ | 0.784(0.7835,0.7843) | 0.807(0.8069,0.8075) |
| $D_4$ | 0.746(0.7453,0.7461) | 0.806(0.8054,0.8060) |
| $D_5$ | 1.155(1.1541,1.1560) | 0.743(0.7425,0.7437) |

In addition, in Fig.3 the power parameters (i.e. $\alpha$ in Eq. (5)) of the power-law distributions to fit the data of out-of-18km displacements are 4.14, 11.36, 8.96, 7.74, and 5.61 respectively. The fitting parameters reveal that the distributions of displacement are schematically similar in different cities, but with different parameters from city to city. Especially, the difference of the power-law tail is significant between cities. In New York City, we observe steep decay of the amount of long-distance taxi trips, while the decay is much gentler in Dalian and Nanjing.

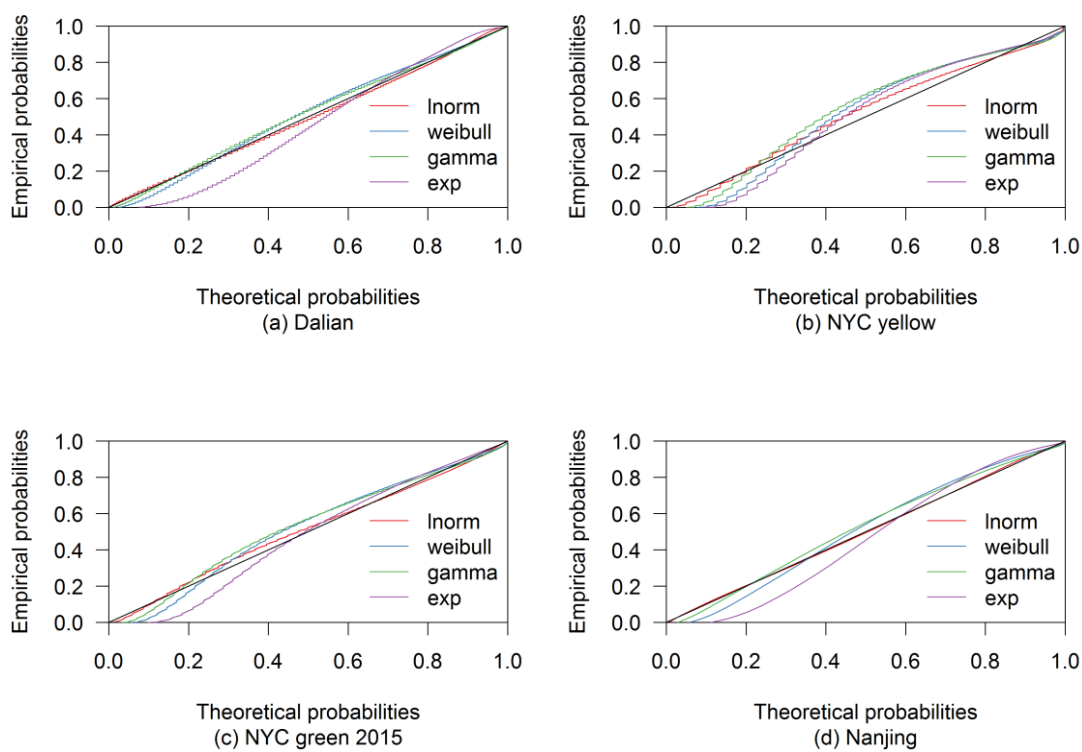

Figure 5 P-P plots of the four candidate models

In order to further illustrate the goodness of fit, the Probability-Probability (P-P) plot [23] of the lognormal, Weibull, gamma and exponential models are drawn in Fig. 5, where theoretical probabilities are on the horizontal axis and empirical probabilities are on the vertical one. When the theoretical probabilities and empirical probabilities are mostly the same, the fitting curve is closer to the diagonal. The P-P plots in Fig. 5 reveal that the displacements in the all examined datasets are best fitted by the lognormal. By comparison, the fittings of the exponential distributions are worse, although some researches have claimed the good fit of this type of distributions [14, 15].

To sum up, the previous investigation reveals that the trips of taxi passengers are comprised of abundant of short-distance trips and relatively-fewer long-distance one. In all the examined datasets, we find the displacements of taxi passenger trips can be best fitted by the lognormal distribution with a power-law tail.

### 4.2 Duration Distribution

Second, we examine the distributions of passenger duration time in the five datasets. The duration time represents the time consumed from the origin to the destination in a passenger's trip. Fig.6 illustrates the box plots of duration time, in which the black dots represent the mean values and the black triangles denote the peak values.

In Fig. 6 we can find that the durations of the five datasets are similar in average, as in all the datasets the mean values are about 15.0 minutes and the median values are about 12.0 minutes. This partly reflects the severer traffic congestion in NYC comparing with Nanjing and Dalian, since the average displacements are shorter in NYC datasets under similar average duration time. Fig. 7 plots the fittings of the five theoretical distributions (i.e. lognormal, Weibull, gamma, exponential and power-law) to the actual duration data series at the log-log scale. Again we skip illustrating the fitting result of the Dataset of "NYC Green 2014" to ease the layout of Fig.7. Similar with the case of displacement distribution, the lognormal model fits the duration data best in all the five datasets.

The goodness-of-fit of the lognormal are also validated by the AIC weights as shown in Table 5.

As same as in the fitting of displacement data, the parameters of the theoretical distributions are computed by MLE. The parameters of the adopted lognormal distribution to fit the duration data with 95% CI are listed in Table 6. Different from the case of displacement distribution, the fitting parameters are quite similar in different cities. This is consistent with the results of Fig. 6.

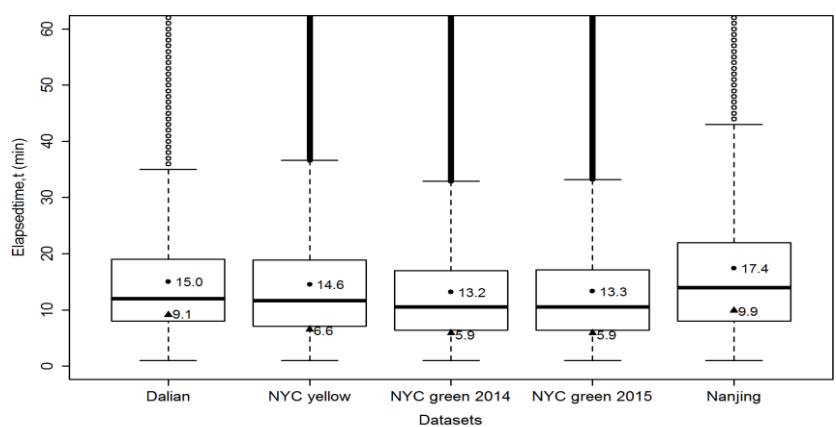

Figure 6 The box plots of duration time in the five datasets

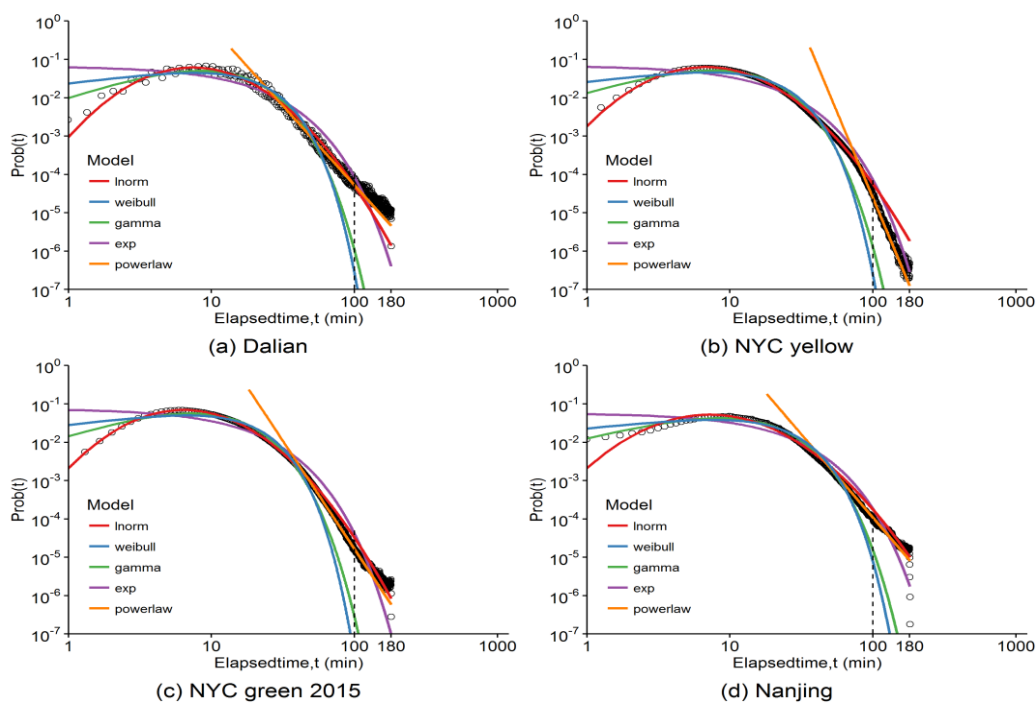

Figure 7 Fittings of the distributions of duration time at log-log scale

In all, the previous analysis shows that the records of durations of trip in all five datasets can best be fitted by the lognormal model, comparing with Weibull, gamma, exponential and power-law models. More accurately, for the trips that elapse more than 100 minutes, the power-law provides best fit. Thus the overall duration records in each dataset are best fitted by the lognormal with a power-law tail. However, more than 99% trips elapse within 60 minutes. Moreover, the fraction of the trips that are more than 100 minutes is at the level of $10^{-4}$ as shown in Fig. 7. Thus the power-law tail is to some extent negligible.

Table 5 The AIC weights of duration time

| Datasets | AIC weight | | | | |
| --- | --- | --- | --- | --- | --- |
| | lognormal | Weibull | gamma | exponential | power-law |
| $D_1$ | 1 | 0 | 0 | 0 | 0 |
| $D_2$ | 1 | 0 | 0 | 0 | 0 |
| $D_3$ | 1 | 0 | 0 | 0 | 0 |
| $D_4$ | 1 | 0 | 0 | 0 | 0 |
| $D_5$ | 1 | 0 | 0 | 0 | 0 |

Table 6 Parameters for the lognormal distributions to fit duration data with 95% CI

| Datasets | MLE for lognormal parameters (with 95% CI bounds) | |
| --- | --- | --- |
| | mean | standard deviation |
| $D_1$ | 2.477(2.4765,2.4770) | 0.691(0.6912,0.6915) |
| $D_2$ | 2.427(2.4266,2.4273) | 0.719(0.7197,0.7191) |
| $D_3$ | 2.344(2.3434,2.3441) | 0.695(0.6950,0.6955) |
| $D_4$ | 2.347(2.3470,2.3476) | 0.702(0.7022,0.7026) |
| $D_5$ | 2.582(2.5812,2.5830) | 0.780(0.7792,0.7804) |

## 4.3 The Distribution of Vacant Time

In previous sub-sections we study the distributions of passenger travel distances and duration time that directly reflect traffic demand. Another interesting issue is the distribution of vacant time of the taxicabs, which reflect traffic demand indirectly. The short vacant time implies the high density of taxi travels.

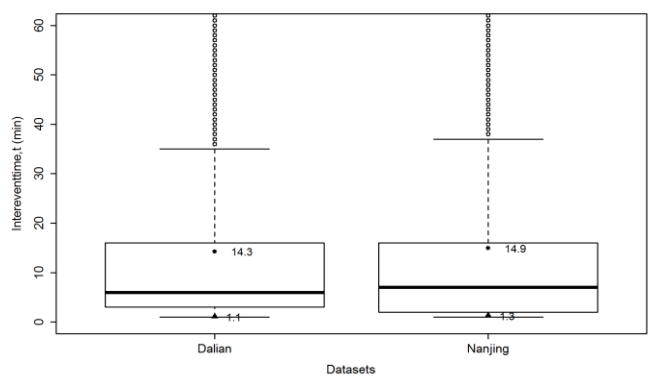

Figure 8 The box plots of vacant time in *D1* and *D5* datasets

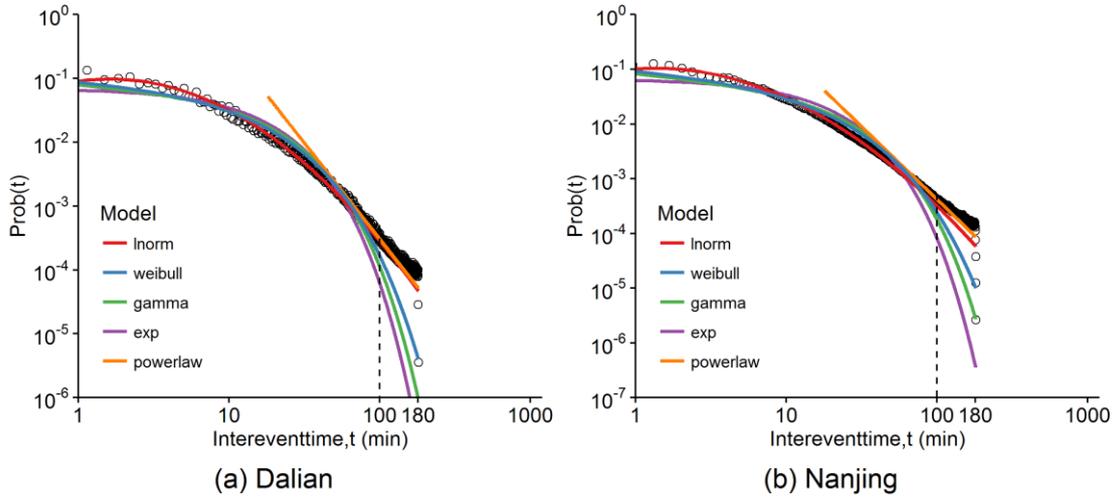

(a) Dalian  (b) Nanjing

Figure 9 Fittings of the distributions of interevent time at log-log scale

Table 7 The AIC weights and parameters for four candidate models of vacant time

| Model | | MLE for parameters (with 95% CI bounds) | | AIC weight |
|---|---|---|---|---|
| | | $D_1$ | $D_5$ | |
| lognormal | mean | 1.942(1.9411,1.9420) | 1.898(1.8965,1.8995) | 1 |
| | sd | 1.205(1.2042,1.2049) | 1.269(1.2676,1.2698) | |
| Weibull | shape | 0.841(0.8412,0.8416) | 0.790(0.7897,0.7911) | 0 |
| | scale | 12.820(12.8142,12.8264) | 12.722(12.7009,12.7421) | |
| gamma | shape | 0.825(0.8248,0.8256) | 0.744(0.7431,0.7453) | 0 |
| | rate | 0.058 (0.0578,0.0579) | 0.050(0.0498,0.0500) | |
| exponential | rate | 0.070 (0.0700,0.0701) | 0.067(0.0670,0.0671) | 0 |

As in the NYC datasets (i.e. $D_2$, $D_3$ & $D_4$) the vehicleIDs are anonymized, the information of vacant time of taxicabs are not available. Therefore we analyze the distributions of vacant time by using the Dalian ($D_1$) and Nanjing ($D_5$) datasets.

Fig. 8 presents the box plots of vacant time in $D_1$ and $D_5$. In both datasets, the mean values are about 14.0 minutes and the median values are about 7.0 minutes. In addition, the peak values are about 1.2 minutes. The box plots also reveal that the distributions of vacant time are right-skewed in the taxi records of both cities.

Fig. 9 plots the fitting results of the four candidate models (i.e. lognormal, Weibull, gamma, and exponential distributions) to the actual data of vacant time, indicating the best fit of the lognormal distribution in both datasets. It is interesting to note that a power-law tail can also be observed, as same as in the distributions of passenger displacements and durations.

Table 7 lists the parameters of the four models to fit the vacant-time data, as well as the AIC weights. The AIC weights of lognormal distribution is 1, while those of the other three models are all zero. This further confirms the goodness-of-fit of the lognormal model comparing with the other three models.

## 5. Discussion and Conclusion

By examining five datasets in three cities, in this paper we find that the displacements and durations of taxi passengers, as well as the vacant time of taxicabs, can best be approximated by the lognormal with a power-law tail. This result can be further discussed in the context of related work.

Our investigation firstly reveals that the power-law distribution, which has often been observed in long-distance travels, does not provide a good fit to the taxi-taking travels within the range of intra-urban areas. It is interesting to note that in Brockmann et al.'s [7] well-noted contribution the power-law fits the actual data well for the travels that are across the distances from $10^1$ km to around $10^3$ km, while the fit is less satisfactory for the travels that are shorter than 10km. By contrast, the main finding of the present work is that the lognormal perfectly fits both the taxi passengers' displacements and durations for the travels that are shorter than 18km in distance and 60 minutes in time, while the longer-distance travels are better fitted by the power-law. Putting the results of the two endeavors together, we conjecture that human mobility may not follow a simple universal scaling-rule. At the scale of global and nation-wide travels, the power-law may be an appropriate approximation, while at the scale of intra-urban areas, the travels may probably be better approximated by a lognormal.

Second we compare our work with the related endeavors on the human mobility patterns in intra-urban areas, especially those endeavors that investigate the statistical characteristics of taxi travels.

The result of displacement distribution obtained in this work basically agrees with that in Csáji et al.'s work [24], which declared a lognormal distribution of travel distances of the mobile-phone users in Portugal. This work provides some support on their result with denser and more accurate spatiotemporal data.

For the research method, our work is similar with Yao et al.'s [12]. However, the results of the two endeavors are quite different, as they suggested the power-law distribution of travel distances of taxi passengers through the analysis of one-day taxi GPS trajectories in one South China city. We argue that our results may be more accurate in that our analysis is based on trajectory data in much longer periods to make the characteristic displacements and durations more prominent. Intuitively, the lognormal approximation is also more reasonable than the power-law approximation for the travels of taxi passengers, since taxi is not the most preferable option for average travelers in their extremely short and long travels. Thus, it would be reasonable to estimate that a characteristic displacement (and duration time) can be observed, while such characteristic value is missing in the power-law distribution. In Yao et al.'s work [12], the power-law distribution of displacement is further explained by a dynamic model that extends the Maxwell-Boltzmann model. This model is also questionable as it assumes human movements are analogous to the random movements of gas molecules. However, the actual movements of citizens are usually purposeful and a great fraction of intra-urban movements are the routine travels, e.g. between residential quarters and business and commercial centers (i.e. the working places and shopping areas). The aggregation of living, working, and shopping and recreational areas in a city often causes a great proportion of travels are converged to some specific distances. This phenomenon is not well-described by a power-law distribution, while the explanation of the normal and lognormal distributions is more reasonable.

The lognormal distributions of the durations of occupied trips and the vacant time of taxicabs identified in this work agree with the corresponding results in Wang et al. [15]. However, for the distributions of displacements, our result is quite different from theirs, as well as Liang et al.'s [14], as both the previous studies have shown the exponential law of displacement distributions. Our examination reveals that the fit of the exponential distribution to actual data of displacements is unsatisfactory for short-distance travels, although it provides good fit to the relatively long travels. In both Wang et al.'s [15] and Liang et al.'s [14] investigations the travels that are shorter than 1km are omitted. As noticed in the Introduction section, some other researches also declare other types of distributions for the displacements of intra-urban travels, e.g. the gamma distribution by examining the taxi data in Lisbon [16], and the negatively binomial distribution according to the investigation of London subway data [17]. The diversity of results may reflect the intricacy of human mobility patterns in intra-urban areas.

Based on the previous analyses and discussions, conclusions of this work are drawn subsequently.

In this work our focus is to examine the pattern of intra-urban travels of taxi passengers. Under such restriction, we find the lognormal distribution with a power-law tail can best fit the displacements of occupied trips, the durations of occupied trips, and the vacant time between two occupied trips. This result may reflect the general mobility pattern of the taxi passengers in intra-urban areas. First, the probability for taking taxi to travel a very-short distance and duration is low. This may be explained by the fact that people are unlikely to take a taxi when the destination is easy to go on foot.

Second, the likeliness for an average passenger to take taxi would steeply increase after the travel distance exceeds the limit of walkable distance. Then the highest probability for passengers to take taxis is reached at a range of distances that are relatively short (e.g. around 1.0km in New York City, 1.7km in Nanjing, and 2.9km in Dalian as shown in Fig.2). Third, the probability gently decays for the taxi trips with longer distance and duration time. When the travel distance is extremely long (e.g. across the boundary of the city), the probability of taxi-taking becomes very low again, as same as in the case of extremely short travel distance.

The previous phenomena of steep increase and relatively-gentle decrease of the probability of taxi trip may possibly ascribe to the combinatorial effect of two factors, the traffic demands itself and the passengers' preference to select taxi as the transport mode. For the general traffic demand, a great proportion of traffics are routinized between passengers' often-visited places (e.g. residential places, the working places, and shopping centers). Furthermore, a great proportion of such routinized trips are across relatively short distances. For example, one shopper tends to select a shopping center that is close to her living place. On the other hand, leveraging the time efficiency and economic cost, the motive for taking-taxi to cross a relatively short distance is usually higher than to cross a long distance. However, the decay of taxi-taking demand is gentle in the reasonable distances. For the travels that cross very long distances, people are more likely to select the substitutional means of transport such as train, subway, and bus; and the amount of taxi trips greatly drops.

To sum up, the exploration in this work sheds some lights on the collective mobility pattern of intra-urban transport by taxi, which is helpful to deepen our understandings on the urban traffic demands in general, and on the taxi demands in particular. Thus the contribution of this work is of potentially practical value in traffic management and urban planning. In addition, the result of this work is of noticeable implication for the theoretical issue of human mobility patterns and statistical rules.

Comparing with the eye-catching power-law or scale-free distributions of travel distances such as in [7], in this work we obtain a somehow "ordinary" result, which accords with the more conventional estimation on the statistical pattern of intra-urban human mobility. The significant difference between the lognormal model in our work and the power-law model suggested by Brockmann et al. [7] and others implies that there may not be a simple answer in the pursuit of the universal scaling-law of human mobility. The patterns of human mobility may probably be influenced by multiple factors, especially by the distance scale of travel and the means of travel (e.g. by airplane, by train, by bus, by taxi, or on foot). If considering the distance scale to more-than-10km, the power-law fit may be appropriate. But the collective travelling data may better be fitted by the lognormal model if the distance scale is squeezed to around 0.5km to

1.0km. At this distance scale, the common mode of mobility is travelling by some automotive vehicle within the boundary of a city or town. Furthermore, if all the on-foot travels can be collected at the even-shorter distance scale, the appropriate theoretical distribution is likely to shift again. Therefore, a noticeable implication of the present work lies in that it indicates that there is probably no "one-size-fits-all" theoretical distribution to perfectly fit human mobility patterns at all distance scales.

As future work, we are going to extend the current research in two directions. First, in the current research, we only take the distance and time of taxi trips into account; and more accurate geographical and time information of the travels is neglected. More thorough spatiotemporal mining of taxi travel records will be carried out, in order to obtain more accurate results. For example, an implication of this work is that the distinction of functional quarters in cities may be a vital cause for the aggregation of taxi-taking trips around some particular distances. We are to examine this by giving further analysis upon the basis of origin-destination data mining [25]. Another direction of future research is to cover different means of transport. In the current work, we only consider the travels by taxi due to lacking the data of other means of transport. Next we are to seek the available data of other means of public transport, especially the travel data of smart transport cards in different cities, and to study the mobility patterns of the passengers who use multiple means of public transport.